\documentclass[twocolumn,showpacs, prd, nofootinbib,superscriptaddress,amsmath,amssymb,floatfix]{revtex4-1}
\usepackage{graphicx,color}  
\usepackage{dcolumn}   
\usepackage{bm}        
\usepackage{amssymb}   
\usepackage{lineno, soul}
\usepackage[dvipsnames]{xcolor}
\usepackage{color}
\usepackage{natbib}
\usepackage[normalem]{ulem}
\usepackage{booktabs,multirow}

\newcommand{\BibitemShut}[1]{}

\def \aap  {A\&A}

\def \apj  {ApJ}

\def \apjl  {ApJL}

\def \apss  {AP\&SS}

\def \prd {Phy. Rev. D}

\def \mnras {MNRAS}

\def \prl {PRL}

\begin{document}

\widetext


\title{Does the Geminga, Monogem and PSR J0622+3749 $\gamma$-ray halos imply slow diffusion around pulsars?} 

\author{S. Recchia}
\affiliation{Department of Physics, University of Torino, via P. Giuria, 1, 10125 Torino, Italy}
\affiliation{Istituto Nazionale di Fisica Nucleare, via P. Giuria, 1, 10125 Torino, Italy}
\author{M. Di Mauro}
\affiliation{Istituto Nazionale di Fisica Nucleare, via P. Giuria, 1, 10125 Torino, Italy}
\author{F. A. Aharonian}
\affiliation{Dublin Institute for Advanced Studies, 31 Fitzwilliam Place, Dublin 2, Ireland}
\affiliation{Max-Planck-Institut für Kernphysik, Postfach 103980, D-69029 Heidelberg, Germany}
\author{L. Orusa}
\affiliation{Department of Physics, University of Torino, via P. Giuria, 1, 10125 Torino, Italy}
\affiliation{Istituto Nazionale di Fisica Nucleare, via P. Giuria, 1, 10125 Torino, Italy}
\author{F. Donato}
\affiliation{Department of Physics, University of Torino, via P. Giuria, 1, 10125 Torino, Italy}
\affiliation{Istituto Nazionale di Fisica Nucleare, via P. Giuria, 1, 10125 Torino, Italy}
\author{S. Gabici}
\affiliation{Universit\'e de Paris, CNRS, Astroparticule et Cosmologie, F-75006 Paris, France}
\author{S. Manconi}
\affiliation{Institute for Theoretical Particle Physics and Cosmology, RWTH Aachen University, Sommerfeldstr.\ 16, 52056 Aachen, Germany}

\date{\today}

\begin{abstract}
The HAWC Collaboration has reported the detection of an extended $\gamma$-ray emission around the Geminga and Monogem pulsars of a few degree extension. Very recently, the LHAASO Collaboration released also the data for an extended $\gamma$-ray emission around the pulsar PSR J0622+3749. This flux can be explained with electrons and positrons injected from these sources and their inverse Compton Scattering on the interstellar radiation fields. So far the size of such $\gamma-$ray halos has been interpreted as the result of the diffusion coefficient around the sources being about two orders
of magnitude smaller than the average in the Galaxy. However, this conclusion is driven by the  assumption that particles propagate diffusively right away after the injection without taking into account the ballistic propagation.
The propagation of cosmic-ray leptons in the proximity of the Geminga, Monogem and PSR J0622+3749 pulsars is examined here considering the transition from the quasi-ballistic, valid for the most recently injected particles, to the diffusive transport regime. For typical interstellar values of the diffusion coefficient, the quasi-ballistic regime dominates the lepton distribution up to distances of a few tens of parsec from the pulsar for particle energies above $\sim 10$ TeV. In this regime the resulting $\gamma-$ray source tends to be rather compact, despite particles travel a long distance. Indeed, for larger values of the diffusion coefficient, particles propagate ballistically up to larger distances with the result of a more point-like $\gamma-$ray source.
When such transition is taken into account, a good fit to the HAWC and LHAASO $\gamma-$ray data around Geminga, Monogem and PSR J0622+3749 is obtained without the need to  invoke a strong suppression of the diffusion coefficient.
\end{abstract}

\maketitle


\section{Introduction}

The HAWC Collaboration reported the detection of  extended (few degrees across the sky) very-high-energy (VHE, above 1 TeV) $\gamma$-ray 
halos around the Geminga and Monogem pulsars  \citep{Abeysekara:2017science}. 
Very recently the LHAASO experiment also reported the detection of an extended $\gamma-$ray halo around the pulsar PSR J0622+3749\citep{LHAASO:2021crt}.
These $\gamma$-ray structures, whose 
existence has been predicted a while ago \citep{Felix-book-2004vhec.book.....A}, 
are the result of 
inverse Compton scattering (ICS) of electrons and positrons ($e^{\pm}$) accelerated at the pulsar's wind termination shock and propagating diffusively in the turbulent interstellar medium (ISM).
The pool of target photons is dominated by the interstellar near to far infrared radiation and the 2.7~K Cosmic Microwave Background (CMB).  

The unexpected outcome of such detection was  the small angular size of the $\gamma$-ray halo surrounding the Geminga, Monogem and PSR J0622+3749 pulsars, leading to the conclusion that the  CR diffusion was inhibited within few tens of pc from the pulsar, and consequently the energy dependent CR diffusion coefficient, $D(E)$, should be smaller, by at least two orders of magnitudes, than the {\it nominal} value used in conventional models of propagation of Galactic CRs  \cite{Abeysekara:2017science}. 
Since then, the suppression of the diffusion coefficient around pulsars has become a 
popular hypothesis \cite{Hooper:2017gtd,Tang:2018wyr,Fang:2018qco,DiMauro:2019yvh,DiMauro:2019hwn, Giacinti-2019nbu}, but so far no convincing theoretical explanation of this effect has been proposed (see e.g.~\cite{Lopez-Coto-Giacinti:2017pbk, Evoli-Morlino-2018PhRvD..98f3017E, Liu-Yan-2019zyj}).
Moreover, a suppression of the diffusion coefficient  in the source region may be problematic for the detection of multi-TeV electrons (which are observed up to 20 TeV).
In fact, if such strong suppression is  a common feature of CR sources (for instance supernova remnants, where a suppression over time-scales of $10^4-10^5$ yrs may be even expected, see e.g.~\cite{Recchia-2021-IN} and references therein), not only pulsars, one may wonder whether multi-TeV electrons can actually leak-out of the small-diffusion region, or whether  they  loose  all  their  energy  within  such  region.
This depends on the extension of such region in the source proximity as compared to the diffusion-loss length of multi-TeV particles. Indeed, as shown below,  the latter reads
$r_d \propto \sqrt{D(E) t_{\rm loss}(E)}$, where $t_{\rm loss} \propto 1/E$ is the loss time scale of leptons, and represents the distance travelled diffusively by particles before loosing their energy. Typically, $D(E) \propto E^{\delta}$ with $\delta < 1$ (and positive, see discussion below), so that $r_d \propto E^{\frac{\delta -1}{2}}$ decreases with the particle energy. Thus, if the extension of the low diffusion zone is larger than the diffusion-loss length of particles of $E \lesssim 1\, {\rm TeV}$, also 20 TeV particles will loose all their energy within such region. 
In general, even without assuming a suppressed diffusion everywhere in the Galactic disk, low diffusion zones may pose a serious concern on the location of the sources of 20 TeV electrons and their possible impact should be carefully investigated.
\\

In this paper, we demonstrate that there is no need to resort  to suppression of the diffusion coefficient 
to explain the angular size of the $\gamma$-ray halo around Geminga, Monogem and PSR J0622+3749. 
Instead, we show for the first time that  the characteristics of such $\gamma$-ray halos are explained by properly accounting for the transition between two propagation regimes, the ballistic and the 
diffusive propagation. 
We focus first on the cases of Geminga and Monogem (HAWC data) and then apply our model also the the case of PSR J0622+3749 (LHAASO data).

Previous studies  of the HAWC data have been performed under the assumption  
of diffusive CR transport 
at any  time after the injection of $e^{\pm}$. 
However, the propagation of electrons deviates from such a simple picture. At the first stage,
determined by the timescale $\tau_c=3D(E)/c^2$ after injection,  electrons with energy $E$  propagate 
ballistically. Then, as time passes, the multiple deflections experienced in the turbulent 
circumstellar magnetic field lead to the isotropization of the particle directions, i.e.  
the propagation proceeds in the diffusive regime. The formal application of the 
diffusion theory  to timescales smaller than $\tau_c$, faces the so-called 
superluminal propagation problem \citep{Aloisio-2005ApJ...625..249A, Parizot-2008A&A...479...97G, Aloisio-2009, Prosekin-2015}. This can be seen by comparing the diffusion, $r_{\rm{diff}}$, and ballistic, $r_{\rm{ball}}$,  distances  travelled  over a time $t=\alpha \tau_c $: 
$r_{\rm{diff}} \sim \sqrt{D\alpha\tau_c} \sim \sqrt{\alpha} \tau_c c$ and $r_{\rm{ball}} =\alpha \tau_c c$, respectively.
For $t \lesssim \tau_c $ ($\alpha < 1 $), $r_{\rm{diff}}>r_{\rm{ball}}$ and the propagation speed in the diffusive regime would  exceed the speed of light.
A fully relativistic extension of the diffusion equation, which would solve such problem, has not been found yet (see e.g.~\cite{Dunkel-2007PhRvD..75d3001D, Aloisio-2009} for a discussion), so that it is necessary to adopt approximate solutions, in which the ballistic and diffusive regimes are recovered in the appropriate limits and the two solutions are somehow sewed  together (see e.g~\cite{Aloisio-2005ApJ...625..249A, Malkov-2017PhRvD..95b3007M} and Eq.~24 of \cite{Prosekin-2015}).

The CR transport is characterized by three regimes depending on the time $t$ after the 
injection: ballistic (for $t<< \tau_c$), diffusive (for $t > \tau_c$) and a transition between the two, that we call quasi-ballistic. 
The transition is governed by the energy-dependent mean free path $\lambda_c(E)$, which, for relativistic particles, is linked to the energy-dependent (as inferred both from theory and from the Galactic CR transport phenomenology \citep{Strong-2007ARNPS..57..285S, Reinert:2017aga, Genolini:2019ewc}) isotropic diffusion coefficient through $D(E) =  \lambda_c(E)\, c/3$ (see e.g.~\cite{Strong-2007ARNPS..57..285S, Subedi-2017ApJ...837..140S}). The time and spatial scales for the isotropization 
are given by $\lambda_c(E)$ and $\tau_c = \lambda_c/c$ respectively (see e.g~\cite{Aloisio-2009, Prosekin-2015, Mertsch-2020Ap&SS.365..135M}). 
In the case of a continuous source, such as pulsars, this also results in the fact that the CR spatial distribution at a distance from the source smaller than $\lambda_c$, is dominated by particles injected within the last $\tau_c$ \citep{Prosekin-2015}. 
Particles emitted at an instant earlier than $ \tau_c$, with respect to the current time, have been isotropized and thus can be treated in the diffusive approximation, while particles injected within the last  $\tau_c$ should be treated in the quasi-ballistic regime. 

The $\gamma$ rays detected by HAWC at energies $5-50$ TeV are mostly produced through ICS predominantly by $e^{\pm}$ of energy  between $20-200$ TeV (see Fig.~2 of \cite{DiMauro:2020jbz}).
Assuming that at these energies the standard diffusion coefficient is $D \gtrsim 10^{29}-10^{30} \, \rm cm^2/s$ \citep{Strong-2007ARNPS..57..285S, Reinert:2017aga, Genolini:2019ewc}, we find that $\lambda_c \gtrsim 3-30 \,\rm pc$. 
Given that  the spatial extension measured for the $\gamma$-ray halo around Geminga and Monogem is $\sim 10$~pc  \citep{Abeysekara:2017science}, the correct treatment of the transition from  ballistic to diffusive propagation  is critical for the interpretation of $\gamma-$ray data.
This is probably the case also for the Fermi-LAT data analysed by \cite{DiMauro:2019yvh}. Indeed, such data refer to $e^{\pm}$ of energy $E\sim 100-1000$ GeV, which correspond to $\lambda_c \gtrsim 3-10$ pc.
However, here we focus on multi-TeV energies  and we postpone the analysis of Fermi-LAT data to a future work.

In this paper, we assume isotropic diffusion,  which typically can be applied,  as a good approximation, to  different   astrophysical environments (see e.g.~\cite{Strong-2007ARNPS..57..285S, Yan-Lazarian-2008ApJ...673..942Y, Subedi-2017ApJ...837..140S} for a discussion). In particular, this approximation has been much used in the modelling of CR propagation around pulsars \citep{Abeysekara:2017science, Hooper:2017gtd, Tang:2018wyr,Fang:2018qco,DiMauro:2019yvh,DiMauro:2019hwn, Giacinti-2019nbu}. 
Note, however, that the applicability of this approach often depends on poorly constrained parameters such as the coherence length of the background magnetic field and its level of turbulence~\citep{Casse-2002PhRvD..65b3002C, Nava-2013MNRAS.429.1643N, Subedi-2017ApJ...837..140S, Lopez-Coto-Giacinti:2017pbk, Giacinti-2019ICRC...36..685G}, and ultimately on the unknown time and space dependent configurations of the field lines in the source region (see e.g~\cite{Nava-2013MNRAS.429.1643N, Shukurov-2017ApJ...839L..16S, Mertsch-2020Ap&SS.365..135M} and references therein). 
A detailed treatment of such issues is beyond the scope of the present work. 


\section{Ballistic-diffusive propagation transition}

We treat the diffusion coefficient as an energy-dependent parameter. The standard values, as deduced from  Galactic CR propagation, are:
\begin{equation}\label{eq:D-E}
    D(E) \approx  D_{0} E_{\rm GeV}^{\delta} \rm \, cm^2/s,
\end{equation}
where $E_{\rm GeV}$ is the particle energy in GeV, $D_{0} \sim 1-4\times 10^{28} \rm  cm^2/s$ and $\delta \sim 0.3-0.6$, with $\delta = 1/3$ corresponding to a Kolmogorov-type turbulence and $\delta=1/2$  to a Kraichnan-type tubulence \citep{Strong-2007ARNPS..57..285S}.
Notice that what matters in the present analysis is the CR diffusion coefficient in the energy range 20-200 TeV. The way in which such coefficient is extrapolated from $\rm 1~GeV$ is not very relevant, but we adopt the typical parametrization of $D$, as in Eq.~\ref{eq:D-E}, in order to frame our results in the context of typical models of Galactic CR propagation.
For the given value of the diffusion coefficient, the mean free path $\lambda_c$ reads
\begin{equation}\label{eq:lc}
    \lambda_c (E_{\rm GeV}) \approx 0.3\, D_{0, 28} E_{\rm GeV}^{\delta} \rm \, pc,
\end{equation}
where $E_{\rm GeV}$ is the particle energy in GeV, $D_{0,28}$ is the diffusion coefficient at 1 GeV in units of $10^{28}\rm cm^2/s$.

\begin{figure}
\centering {
\includegraphics[width=0.50\textwidth]{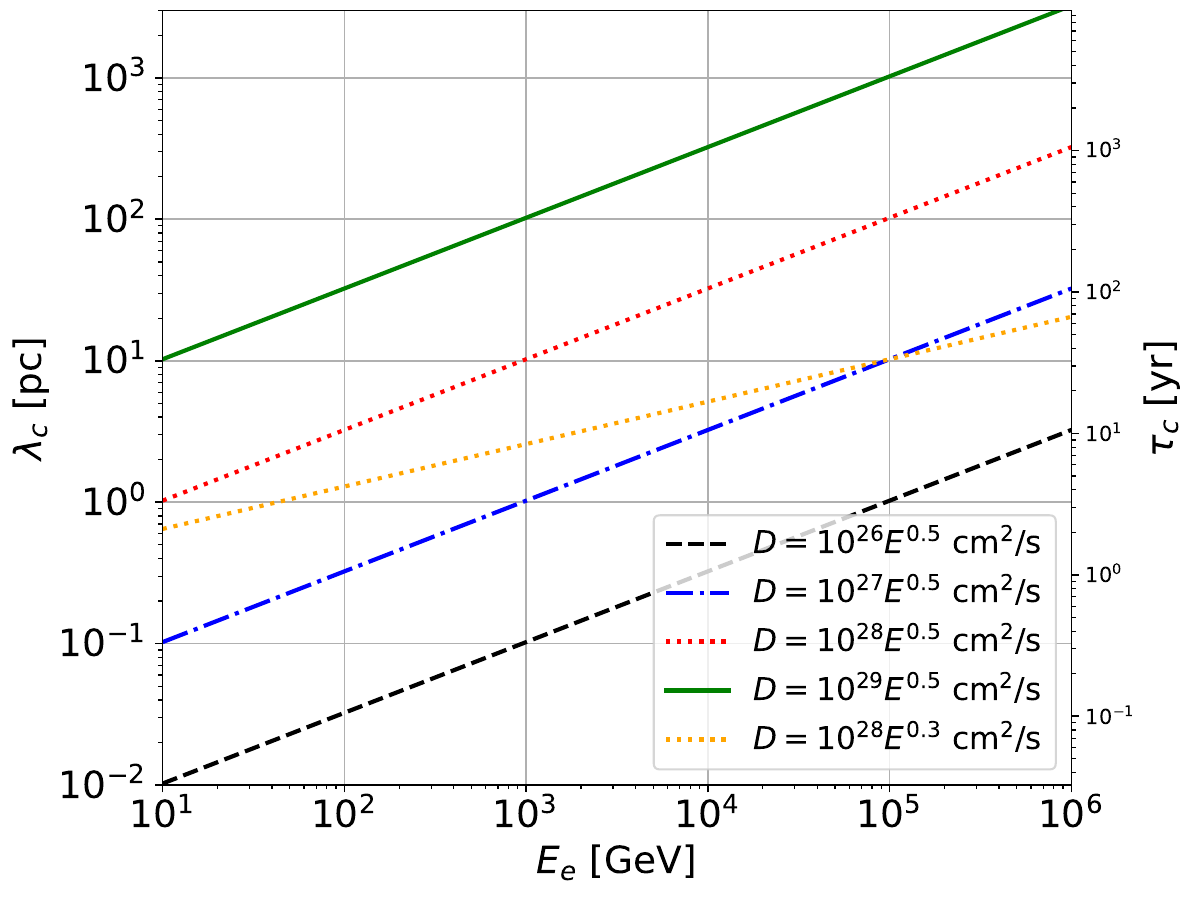}
}
\caption{
Mean free path, $\lambda_c$, and scattering time-scale $\tau_c$, as a function of the particle energy. The curves refer to different values of the normalization of the diffusion coefficient, $D_0$, and of the slope, assumed as $
\delta = 0.5$ or 0.3 as marked.
}
\label{fig:lc-tc}
\end{figure}

Unless otherwise stated, we assume that $\delta=0.5$ which is compatible with recent analysis of CR data (see e.g., \cite{Genolini:2019ewc,DiMauro:2021qcf}) and with the results of simulations for diffusive propagation \cite{Reichherzer:2021yyd}. Previous papers have assumed a value of $1/3$ \cite{Abeysekara:2017science,DiMauro:2019hwn,DiMauro:2019yvh,Tang:2018wyr}. We will explain later on how the results are modified with $\delta=1/3$.
The mean free path increases with the particle energy and with the overall normalization $D_{0}$. For example, at 100 GeV (10 TeV) it is 3 (30) pc for $D_0= 10^{28}$ cm$^2$/s and 0.03 (0.3) pc for $D_0= 10^{26}$ cm$^2$/s. As a consequence, the quasi-ballistic propagation is relevant up to larger distances from the source  for increasing particle energy and larger $D_0$.
This can be seen in Fig.~\ref{fig:lc-tc}, where $\lambda_c$ and $\tau_c$ are shown as a function of the particle energy, for several values of $D_0$ and for $\delta = 0.5$ and 0.3.

We assume that a pulsar of age $T$ turns on at $t =0$ and injects leptons following the time dependent spin down luminosity $L(t)$:
\begin{align}\label{eq:Lt}
     L(t) = \eta L_0 \left(1+\frac{t}{\tau_0} \right)^{(-(n+1)/(n-1))},
\end{align}
where $L_0$ is the initial spin-down luminosity, $n$ is the braking index (assumed to be 3) and $\tau_0$ is the typical pulsar spin down timescale, which we take equal to 12~kyr as in  \cite{Abeysekara:2017science,DiMauro:2019hwn,DiMauro:2019yvh,Tang:2018wyr}. \\

In the diffusive regime (particles injected at times $t_0 \leq  T-\tau_c$), the electron density, $f_{\rm{diff}}$,  at a distance $r$ from the source of age $T$, taking into account diffusion and energy losses is, under the assumption of  continuous injection (see, e.g.~\cite{DiMauro:2019yvh}), given by:
\begin{align}\label{eq:fe-full}
     f_{\rm{diff}}(r, E) & = \int_0^{T -\tau_c}  d t_0
     \frac{Q(E_0)L(t_0)}{\pi^{3/2}r_d^3(E,E_0)} \frac{b(E_0)}{b(E)} e^{-\frac{r^2}{r_d^2(E, E_0)}},
\end{align}
where $b(E) = dE/dt$ is the energy loss rate. This is computed including a fully relativistic calculation of the ICS losses (see \cite{DiMauro:2020cbn} for  details) using the interstellar radiation field model as in \cite{Vernetto:2016alq} and the synchrotron radiation losses assuming a Galactic magnetic field of $3\,\mu$G, which are the relevant loss mechanisms for multi-TeV $e^{\pm}$. Particles emitted from the source at time $t_0$ with energy $E_0$ cool down to energy $E$ during the time $T-t_0$. The conclusions of our letter do not change if we use a different model for the ISRF, as in \cite{Porter_2006}, or if we vary the strength of the Galactic magnetic field around $3\,\mu$G. 
$r_d$ is the propagation length due to diffusion and energy losses and it is defined as:
\begin{equation}\label{eq:rd}
r^2_d (E,E_0) = 4 \int^{E_0}_E D(E')/b(E') dE'.
\end{equation}
$Q(E) \propto E^{-\alpha}\,e^{-E/E_c}$ is the injection spectrum, that here we take as a power-law with index $\alpha = 1.5$  and a cutoff at an energy of $E_c = 150$  TeV. We fix the cutoff energy because a softening of the spectrum is needed to fit well the HAWC spectrum of Geminga in previous analyses \cite{Abeysekara:2017science,DiMauro:2019yvh} and the LHAASO source PSR J0622+3749\citep{LHAASO:2021crt}.
The spectral shape for the $e^{\pm}$ injected spectrum is compatible with multiwavelenght observations of PWNe even if the values of the parameters are not well constrained \cite{Torres:2014iua}.
The injection spectrum is normalized as:
\begin{align}\label{eq:fball}
\int_0^T dt\int_{0.1\rm{GeV}}^{\infty} E Q(E, t) dE = \eta W_0,
\end{align}
where $W_0$ is the total spin down energy \cite{DiMauro:2019yvh,DiMauro:2019hwn} and $Q(E,t)$ is given by $Q(E,t) = Q(E)\cdot L(t)$.

In the ballistic regime (particles injected at times $ T-\tau_c < t_0 \leq T$), the  $e^{\pm}$ density, $f_{ball}$, is given by \citep{Aloisio-2005ApJ...625..249A, Prosekin-2015}:
\begin{align}\label{eq:fball}
     f_{ball}(r, E) & = \int_{T -\tau_c}^{T} \frac{Q(E)L(T)}{4\pi c^3 (T-t_0)^2}\delta\left((T-t_0) - \frac{r}{c} \right)dt_0 = \\ \nonumber 
    & =\frac{Q(E)L(T)}{4\pi c r^2}H(\tau_c\, c -r),
\end{align}
where $H(\tau_c\,c -r)$ is the Heaviside function, which is zero for $r>\tau_c\,c$. Since the typical spin-down time-scale $\tau_0$ (few kyrs) is much larger than $\tau_c$ (at most few tens yrs) we assume that the luminosity is constant 
and equal to $L(T)$. We also neglect energy losses since  for energies of about 10 TeV and $D_0\sim 10^{28}$ cm$^2$/s, $\tau_c \sim 100$ years and thus much shorter than the energy loss time.
In order to have a smooth transition between the two regimes, we substitute the term $H(\tau_c\,c -r)$ in Eq.~\ref{eq:fball} with $\exp{[ - (r/(2\lambda_c))^2 ]}$. We find that our results are similar by choosing other smoothing functions, such as $\exp{[ - r/(2\lambda_c) ]}$.
The total $e^{\pm}$ density is given by $f_{e}(r, E) = f_{\rm{ball}}(r, E) + f_{\rm{diff}}(r, E)$.

Given the relativistic nature of the  ICS, $\gamma$ rays are mainly emitted along the direction of the momentum  of the parent CR. When the CR distribution is isotropic (diffusive regime),  at any location around the source there will be CRs traveling in each direction, and the size of the $\gamma-$ray halo reflects the size of the  $e^{\pm}$ halo. Instead, in the purely ballistic regime the CR angular distribution is strictly anisotropic, which would lead to a point $\gamma-$ray source, given by the $\gamma-$rays produced by $e^{\pm}$ that point toward us \citep{Gabici-2005PhRvL..95y1102G,2010PhRvD..82d3002A,Prosekin-2015}, no matter the extension of the $e^{\pm}$ halo. 
In the quasi-ballistic regime the $\gamma-$ray halo size is intermediate between the two cases.
The angular distribution of particles in the transition from the ballistic to the diffusive regime can be calculated 
in the small-angle diffusion approximation with the following distribution \cite{Prosekin-2015}:
\begin{equation}\label{eq:Mmu}
     M(\mu) = \frac{1}{Z(x)} \exp{\left(-\frac{3(1-\mu)}{x} \right)},
\end{equation}
where $Z(x) = \frac{x}{3}\left(1- \exp(-6/x) \right)$, $x(E) = r c /D(E) = 3\,r /\lambda_c$,  $\mu=(l \cos(\theta)-s)/r$ and $r(s, \theta) = \sqrt{l^2 + s^2 -2\,l\,s\,\cos\theta}$. Here $s$ is the distance along the line of sight, $\theta$ the angle between the source and the line of sight, $l$  the distance from the source and $\mu$ the cosine of the angle between the radial direction and the direction of the line of sight. 
The total  particle distribution function is then given by:
\begin{equation}\label{eq:Fe-full}
    F_e(E, s, \theta) =2 f_{e}(E, r(s, \theta))M(\mu(s,\theta)).
\end{equation}
$M(\mu)$ is normalized as $\int_{-1}^{1} M(\mu) d\,\mu \, =1$.
In the limit $r>> \lambda_c$, i.e. for distances from the source much larger than the mean free path, $M(\mu)$ reduces to $1/2$ (particles pitch angle  uniformly distributed between $-1$ and $1$ around the radial direction) and the total distribution function reduces to the diffusive (isotropic) particle density. 
For $r < \lambda_c$,  $M(\mu)$ encompasses the small angle approximation for the particle propagation and the anisotropic angular distribution of particles in the quasi-ballistic regime. \\  
\begin{figure}
\centering {
\includegraphics[width=0.50\textwidth]{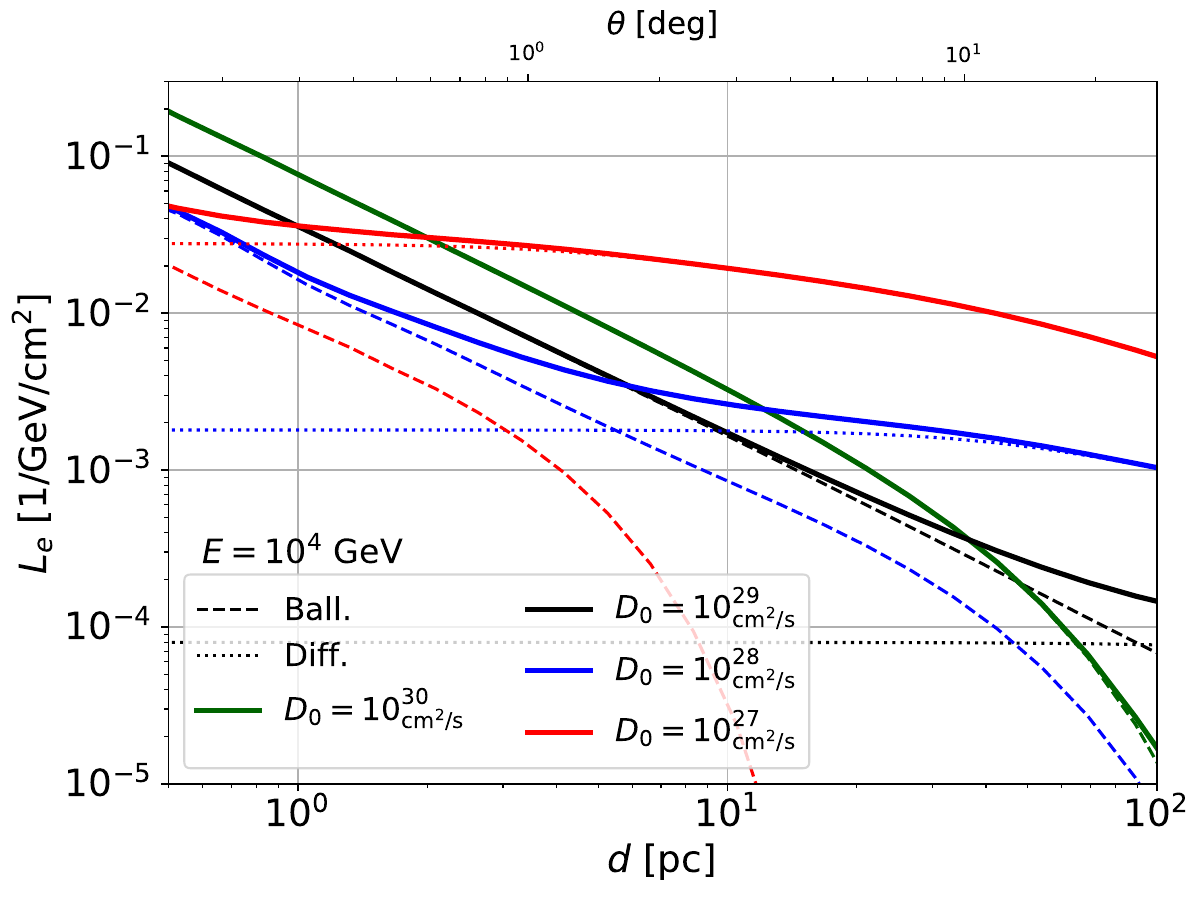}
}
\includegraphics[width=0.50\textwidth]{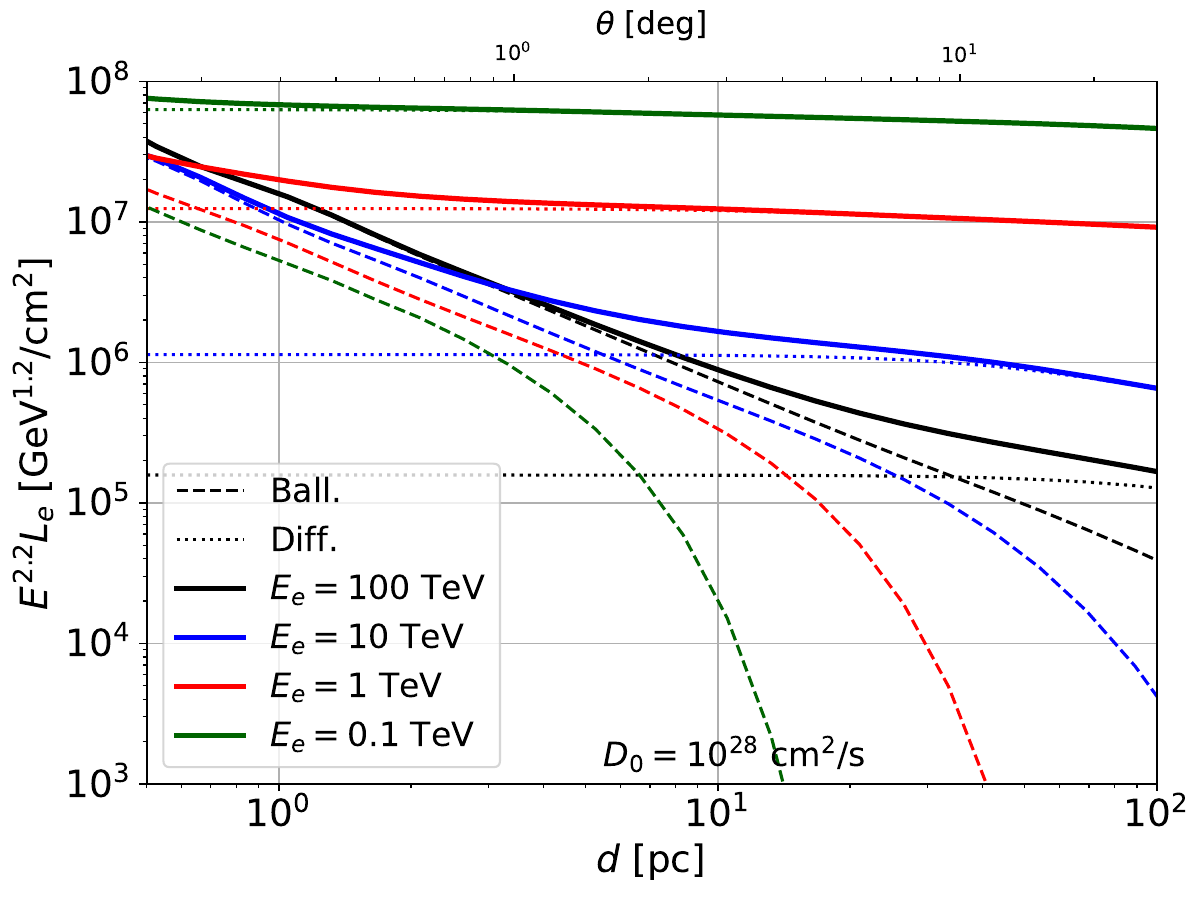}
\caption{Electron density of the Geminga pulsar integrated along the line of sight as a function of the projected distance form the source $d$: \textit{top panel} at 10 TeV and for different values of $D_0$, from $10^{27}\, \rm cm^2/s$ to $10^{30}\, \rm cm^2/s$; \textit{bottom panel} at different energies from 100 GeV to 100 TeV for $D_0 = 10^{28}$ cm$^2$/s. The dashed lines correspond to the quasi-ballistic regime contribution, the dotted line to the diffusive contribution and the continuous lines to the total.}
\label{fig:D1e28-Le-test}
\end{figure}

In order to show the contribution of the ballistic and diffusive regime to the $\gamma-$ray emission we integrate the electron distribution function $F_e$ along $s$ \citep{Berezinskii-1991AN....312..413B}:
\begin{align}\label{eq:Le-full}
     L_e(E,\theta) = \int^{\infty}_{0} ds F_{e}(E, s,\theta).
\end{align}
$L_e(E,\theta)$ reflects  the spatial profile of the  $\gamma-$ray emission at a given $e^{\pm}$ energy as a function of the projected distance from the source $d= l\tan \theta$.\\
\begin{table}
\centering
\begin{tabular}{|c|c|c|c|c|c|}
\hline
& $\dot{E}\, \rm  [erg/s]$ & $T\,\rm [kyr]$
& $l\, \rm [kpc]$ 
& $\tau_0 \,\rm [kyr]$
& $n$ \\
\hline
 Geminga 
& $3.25\times 10^{34}$ & $342$
& $0.19 $ 
& $12.0$
& $3$ \\
\hline
 Monogem 
& $3.8\times 10^{34}$ & $111$
& $0.288 $ 
& $12.0$
& $3$ \\
\hline
PSR J0622+3749
& $2.7\times 10^{34}$ & $208$
& $1.6 $ 
& $12.0$
& $3$ \\
\hline
\multirow{2}{*}{Injection Spectrum} &  \multicolumn{3}{c}{spectral index} & \multicolumn{2}{|c|}{$E_c$}  \\
 \cline{2-6}
  & \multicolumn{3}{c}{1.5} & \multicolumn{2}{|c|}{150 TeV}  \\
 \hline
\end{tabular}
\caption{
Characteristics of Geminga, Monogem and PSR J0622+3749 and injection spectrum parameters: $\dot{E}$ is the current spin-down luminosity, $T$ is the age, $l$ the distance from Earth, $\tau_0$ the assumed spin-down timescale and $n$ the braking index, $E_c$ the cut-off energy.
}
\label{table:pulsar-param}
\end{table}

We perform this calculation for the Geminga pulsar using for the age $T=342$ kyr, the distance $l = 0.19$ kpc and the current spin down luminosity $3.25\times 10^{34}$ erg/s \cite{2017ApJ...835...29Y}.
Instead, for the Monogem pulsar we choose in this paper $T=111$ kyr, $l=0.288$ kpc and $\dot{E}=3.8\cdot 10^{34}$ erg/s using the results of the ATNF catalog\footnote{\url{https://www.atnf.csiro.au/research/pulsar/psrcat/}}.
For PSR J0622+3749 we use $T=208$ kyr, $l=1.6$ kpc and $\dot{E}=2.7\cdot 10^{34}$ \citep{LHAASO:2021crt}. 
The pulsars characteristics are also reported in Tab.~\ref{table:pulsar-param}.
Finally, we fix the efficiency $\eta$ for the conversion of spin-down pulsar luminosity into $e^{\pm}$ to 100\%.
In the top panel of Fig.~\ref{fig:D1e28-Le-test} we fix $E=10$ TeV and change $D_0$ in the range $10^{27}-10^{30}\, \rm cm^2/s $. In the bottom panel we fix $D_0$ to $10^{28}\, \rm cm^2/s $ and change the energy from $0.1-100$ TeV.
We show the results only for $r>0.5$ pc because the bow shock has a similar size (see, e.g., \cite{Caraveo1345}) and so for smaller distances our model might not apply.

All the figures share a general trend.
Up to distances $\approx\lambda_c(E)/3$ from the  pulsar, the most important contribution to $L_e$ comes from $e^{\pm}$ injected most recently, within the last $\tau_c$, that move quasi-ballistically. This gives a spatial profile of $L_e(E, \theta)$ a bit steeper than $\propto 1/r $, expected as due to the angular distribution $M(\mu)$ (see Eq.~\ref{eq:Mmu}) in the quasi-ballistic regime \citep{Prosekin-2015}. At $d \gtrsim \lambda_c(E)/3$ the main contribution to $L_e$ is due to particles injected at $t_0 < T-\tau_c$, which move diffusively and give a rather flat $L_e$ profile, at least up to a distance $d \approx r_d \sim   \sqrt{4\,D(E)\,t_{\rm{loss}}(E)}$, where $t_{\rm{loss}}$ is the time scale for energy losses. Indeed, at  distances larger than $r_d$ the exponential cut-off term  $\exp(-r^2/r_d^2)$ in the diffusive solution (see Eq.~\ref{eq:fe-full}) becomes more relevant and $L_e$ decreases with $r$ \citep{Recchia-2019PhRvD..99j3022R}.

In the top panel of Fig.~\ref{fig:D1e28-Le-test} we see that for $D_0 = 10^{27}\, \rm cm^2/s$ the ballistic-diffusion transition would happen so close to the  pulsar, for $r < 0.05$ pc and $\theta<0.1^{\circ}$, that the $\gamma-$ray morphology observed by HAWC would be solely determined by the diffusive propagation regime. Instead, for $D_0= 10^{28}-10^{29}\, \rm cm^2/s$, $\lambda_c > 10$ pc and the spatial extension of the $\gamma-$ray halo is mainly  determined by the quasi-ballistic propagation regime, that dominates, for the $D_0= 10^{28}\, \rm cm^2/s$ case, within  an angle $\theta\sim 3^{\circ}-4^{\circ}$. 
We also show that for much larger diffusion coefficient values, i.e.~$D_0\sim10^{29}$ cm$^2$/s, the quasi-ballistic model exhibits a cutoff for distances $d>20$ pc from the  pulsar. Indeed, for such large values of $D_0$, $e^{\pm}$ are moving almost ballistically for hundreds of parsec and thus the $\gamma$-ray source would look as point-like.
In the bottom panel of Fig.~\ref{fig:D1e28-Le-test} we show that for $D_0= 10^{28}\, \rm cm^2/s$ and 100 TeV (10 TeV) the transition happens at a  distance from the  pulsar of $\sim 20$ (5) pc and $\theta\sim 5^{\circ}$ ($1^{\circ}$). 

As a final remark, notice that, at a given $D_0$, changing $\delta$ from 0.5 to 0.3 makes $\lambda_c$ smaller at multi-TeV energies, as shown in  Fig.~\ref{fig:lc-tc}. This would translate in a ballistic-diffusion
transition happening at distances closer to the pulsar.

\section{Fit to the HAWC data for Geminga and Monogem and LHAASO data for PSR J0622+3749}
\label{sec:fit}
Here we perform a fit to the HAWC data for the surface brightness of Geminga and Monogem and   to the LHAASO data for PSR J0622+3749 by using the model that includes both the diffusive and ballistic contribution as reported in the previous section. We take the value of the diffusion coefficient $D_0$ and the $e^{\pm}$ conversion efficiency of the pulsar as free parameters. The efficiency $\eta$ is calculated integrating the pulsar source term above 0.1 GeV as in \cite{DiMauro:2019hwn,DiMauro:2019yvh}.
We also test variations of our benchmark model for Geminga by running the analysis with a distance of 0.25 kpc, obtained with the model as in \cite{Cordes:2002wz}, and trying other two values for the Galactic magnetic field $B$, of 2 to 4 $\mu$G\footnote{The magnetic field is the most relevant quantity for calculating the $e^{\pm}$ energy losses since at energies larger than 10 TeV are dominated by Synchrotron radiation.}.
We do not test any variation for the model applied to Monogem because the data are much less precise than the ones for Geminga and the best fit for the diffusion coefficient is much more uncertain.

\begin{figure}
\centering
\includegraphics[width=1.0\columnwidth]{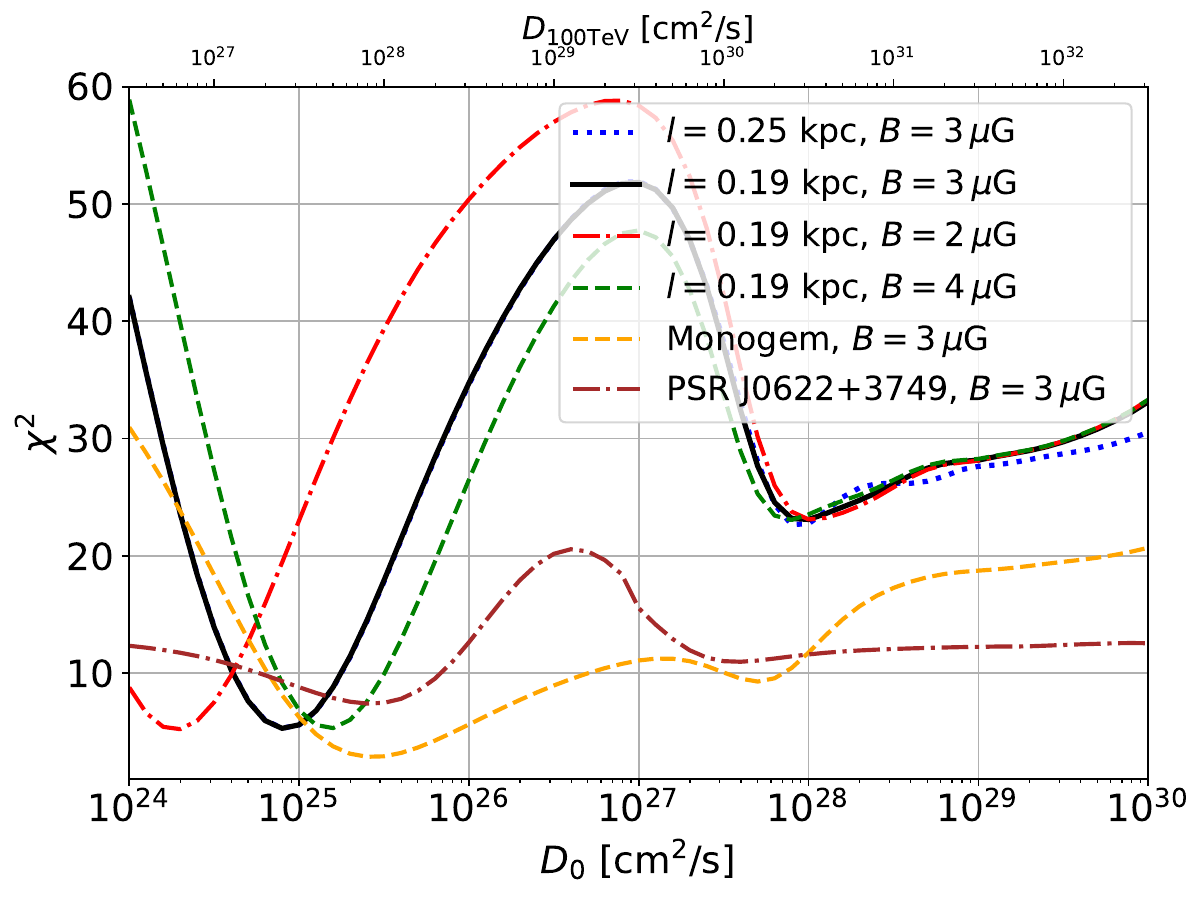}
\caption{
Fit to the HAWC data for Geminga and Monogem, and to the LHAASO data for PSR  J0622+3749: $\chi^2$ as a function of the value of the diffusion coefficient at 1 GeV, $D_0$. We report the results obtained for two different distances of the Geminga pulsar and for different magnetic field values $B$. Instead, for the fit to the Monogem and  PSR J0622+3749 data we only show the data for our benchmark model for the source distance and magnetic field. On the top axis the corresponding value of the diffusion coefficient at 100 TeV is reported.}
\label{fig:Hawc-gemingachi}
\end{figure}

We show the profile of the $\chi^2$ as a function of $D_0$ in Fig.~\ref{fig:Hawc-gemingachi}. The $\chi^2$ for Geminga has a first minimum at about $D_0\sim 0.2-2\times 10^{25}$ cm$^2$/s ($D(100\, {\rm TeV})\sim 0.6-6.4\times 10^{27}\, \rm cm^2/s$) with a best-fit $\chi^2$ of about 5. The best-fit value of $D_0$ decreases with smaller values of $B$, since the last imply a smaller energy loss rate. These values of $D_0$ are at least three orders of magnitude smaller than the results obtained by fitting CR data \cite{Genolini:2019ewc,DiMauro:2021qcf}. 
For larger diffusion coefficients the $\chi^2$ increases and then decreases again, giving a second minimum at about $D_0\sim 0.7-2\times 10^{28}$ cm$^2$/s ($D(100\, {\rm TeV})\sim 2.2-6.4\times 10^{30}\, \rm cm^2/s$) with a best-fit $\chi^2$ of about 22.
This second minimum corresponds to a scenario where the effects of the transition between ballistic and diffusive regimes cannot be ignored,
with the quasi-ballistic propagation dominating  at distances smaller than a few tens of parsec from the source. For larger values of $D_0$ the $\chi^2$ grows slowly because in the quasi-ballistic regime the radial profile is not much influenced by the specific value of $D_0$. 

On the other hand, the $\chi^2$ distribution for Monogem presents a similar behaviour but with much less pronounced variations. There is still a first minimum at about $2-3\cdot 10^{25}$ cm$^2$/s and a second minimum at about $4-8\cdot 10^{27}$ cm$^2$/s.

For the source PSR J0622+3849 we apply the convolution with the PSF of LHAASO since the first two data points of the surface brightness are at an angular distance from the center of the source of about $0.1^{\circ}-0.3^{\circ}$ and the size of the LHAASO PSF is of a similar size.
We convolute our predictions with the PSF of LHAASO:
\begin{equation}
 \frac{d\Phi}{d\theta}_{conv} (\theta)=\int_{-\inf}^{+\inf} \frac{d\Phi}{d\theta'} (\theta') {\rm PSF}(\theta-\theta') d\theta'
\end{equation}
where the PSF is a gaussian with $\sigma$=0.38° (see \cite{LHAASO:2021zta})
The convolution of the surface brightness, if the ballistic regime dominates the morphology, spread the emission from an almost point-like emission to a morphology very similar to the PSF Gaussian.
The result for the $\chi^2$ as a function of $D_0$ are reported in Fig.~\ref{fig:Hawc-gemingachi} where we can conclude that the purely diffusive and the ballistic scenarios provide almost equally good fit to the data with values of the diffusion coefficients of about $1.5-4.0\times 10^{25}$ cm$^2$/s ($D(100\, {\rm TeV})\sim 4.8-12.8\times 10^{27}\, \rm cm^2/s$) and $0.2-2.0\times 10^{28}$ cm$^2$/s ($D(100\, {\rm TeV})\sim 0.6-6.4\times 10^{30}\, \rm cm^2/s$), respectively.

Remarkably, the best-fit values we obtain for the quasi-ballistic scenario in the analysis of all three sources is close to the values found by fitting CR data \cite{Genolini:2019ewc,DiMauro:2021qcf}.
The goodness of the fit for the ballistic case with $\chi^2 \sim 22$ for Geminga and $\sim 9$ for Monogem tells us that this second minimum gives a satisfactory fit to the data,  as the first minimum associated with a suppressed diffusion coefficient. Moreover, such $\chi^2$ values for the quasi-ballistic regime are probably overestimated. Indeed, if we consider Geminga (but similar considerations apply also to Monogem) the $\chi^2$ is mainly dominated by the first data point at $\approx 0.32^{\circ}$, which corresponds to a distance from the source of of 1.1-1.4 pc, depending on the distance of Geminga from us (190-250 pc). 
In fact, if we perform the fit without the first point we obtain for $D_0=10^{28}$ cm$^2$/s a $\chi^2=6$, which is much smaller than the value of 22 we obtain with the entire data set.
At such small distances two effects are relevant. First of all the actual size of the bow shock, which is not precisely known, but may be a sizable fraction of a pc \citep{Caraveo1345}. Obviously our model is not supposed to work within the bow-shock, but only after particles have escaped. Increasing the bow-shock size from 0.3 pc, as we assume, to 1 pc would have the consequence of flattening our predictions in the first data point. 
Second, given the rather sharp profile predicted in the quasi-ballistic regime, the convolution with the PSF of the experiment ($\approx 0.5^{\circ}$ \cite{Abeysekara:2017mjj}) is very important, because it would flatten our predicted radial profile close to the source, making it more compatible to the first data point. In the purely diffusive model the radial profile is already rather flat close to the source and the  convolution with the PSF is not that important. However, it is not possible to properly take into account such effect without the analysis tools of the HAWC collaboration. We postpone to future papers for a more precise treatment of this effect.
Thus, also taking into account these two caveats, both a suppressed diffusion and a typical diffusion scenario give a satisfactory fit to the Geminga and Monogem $\gamma-$ray data,  
and the current precision of the data does not allow to state that statistically one of the two interpretation is preferred. 
On the other hand, if a typical
Galactic diffusion coefficient can account for the data, there is no need to invoke a suppression of diffusion that is difficult to explain theoretically.

When we use a value for the slope of the diffusion coefficient of $\delta = 1/3$ we find that the goodness of fit is similar to the case with $\delta = 0.5$ and the best-fit value for $D_0$ scales, as expected, by the diffusion coefficient dependence with the energy becoming $5\times 10^{25}$ cm$^2$/s ($D(100\, {\rm TeV})\sim 2.3\times 10^{27}\, \rm cm^2/s$) for the diffusive propagation and $5\times 10^{28}$ cm$^2$/s ($D(100\, {\rm TeV})\sim 2.3\times 10^{30}\, \rm cm^2/s$) for the ballistic scenario.\\

The best-fit efficiency obtained for Geminga is between $3-5\%$ in the low diffusion coefficient scenario and $180-200\%$ for the ballistic one, testing different strength for $B$ and the two values of the source distance.
In our benchmark case with $B=3$ $\mu$B the efficiency for the diffusive and quasi-ballistic cases are $(3.8\pm0.4)\%$ and $(190\pm 20)\%$, respectively.
Instead, for Monogem the efficiency in the pure diffusion scenario  is about $2-6\%$ while in the ballistic case is about $60-100\%$. 
The former value  is consistent with the ones some of us obtained in \cite{DiMauro:2019yvh}, if $D_0$ is properly rescaled assuming the different value of $\delta$ used.
Finally, for the source PSR J0622+3849 the efficiency is about $6-10\%$ in case of pure diffusion and about $40-100\%$ in the ballistic case.

The different efficiency value obtained for the purely diffusive and ballistic-diffusive cases is due to the fact (if we assume the same luminosity) that the first produces a flatter  $L_e$  at small distances from the source, while in the quasi-ballistic regime $L_e \sim 1/r$ is  steeper at the same distances (see Fig.~\ref{fig:D1e28-Le-test}). 
As a consequence $L_e$ is much larger for the diffusive case at the angles measured by HAWC ($\theta>0.5^{\circ}$) and so the efficiency must be smaller than the value obtained for the quasi ballistic case.

The spectral shape of the $e^{\pm}$ injected from PWNe is not well known, and the best-fit value of the efficiency can change for different spectral parameters. In particular changing the slope of the injection spectrum to $\gamma_e=1.0$ the efficiency for Geminga in the quasi ballistic case becomes $140\%$. Also the energy range of $e^{\pm}$ injected by pulsars in not precisely known. A lower limit for the efficiency can be found by assuming that the pulsar injects electrons and positron only at energies of interest for the HAWC data. In particular, $\gamma$ rays between $5-50$ TeV are produced mostly by $e^{\pm}$ between $30-300$ TeV. Assuming this range for the energy we find an efficiency of about $90\%$.
There are other effects that might change the value of the efficiency.
For example, the way of treating mathematically the transition between ballistic and diffusive regimes (see \cite{Prosekin-2015} for a discussion) and the form of the function $M(\mu)$ can lead to 
slightly different best-fit values for the diffusion coefficient (which affects the required efficiency). Moreover,  the value of $\tau_0$, as well as  the pulsar spin-down luminosity  $\dot{E}$ and its distance are uncertain. Given the uncertainty of all the above cited parameters the efficiency obtained to fit the HAWC data is not incompatible with the pulsar energetics. Future more precise data by HAWC and LHAASO and at GeV energies by {\it Fermi}-LAT will help to constrain better the injection spectrum of Geminga and in turn test more precisely the energetics needed for this source in the ballistic scenario.

It is also important to remark that, similar to previous analyses of pulsars' halos, we are assuming that particles escape isotropically from the source and undergo  isotropic diffusion. Such setup is the one that tends to maximize the required efficiency, since particles are spread spherically.
However this description may break down within a few pc from the source, where the  the magnetic flux tube that encompasses the source should be taken into account. The resulting propagation topology may then be different from a isotropic diffusion, with a progressive transition from 1-dimensional anisotropic to a 3-dimensional isotropic propagation with the distance from the source (see e.g.~\cite{Nava-2013MNRAS.429.1643N}). In this case the volume occupied by the particles would be smaller, which reduces the required efficiency. A detailed  treatment of these issues is postponed to a forthcoming paper.

A reduction of the required efficiency compared to the isotropic setup could be envisaged if relativistic $e^{\pm}$ are emitted at the pulsar's wind termination shocks in collimated jets. In this case there would be an enhancement of the observed luminosity with respect to the intrinsic one due to the beaming effect, similar to what happens for active galactic nuclei. This effect, if the beams are emitted close to the direction of the line of sight that points to the source, can enhance the luminosity by a factor from a few to a factor of tens (see e.g.~\cite{Dondi-Ghisellini-1995MNRAS.273..583D}).

Such a high efficiency  
agrees well with the PWN paradigm in which a major fraction of the spin-down luminosity of the pulsar is transferred to multi-TeV electrons trough production and termination of the cold ultrarelativistic $e^\pm$ wind \cite{1974MNRAS.167....1R,1984ApJ...283..710K} and less than $10\%$ is transferred to gravitational waves \cite{Abbott-2008ApJ...683L..45A} and protons \cite{2011MNRAS.410..381B}.
In particular, in the case of the Crab Nebula, $\eta$ is very close to $50 \%$.

\begin{figure*}
\centering
\includegraphics[width=1.0\columnwidth]{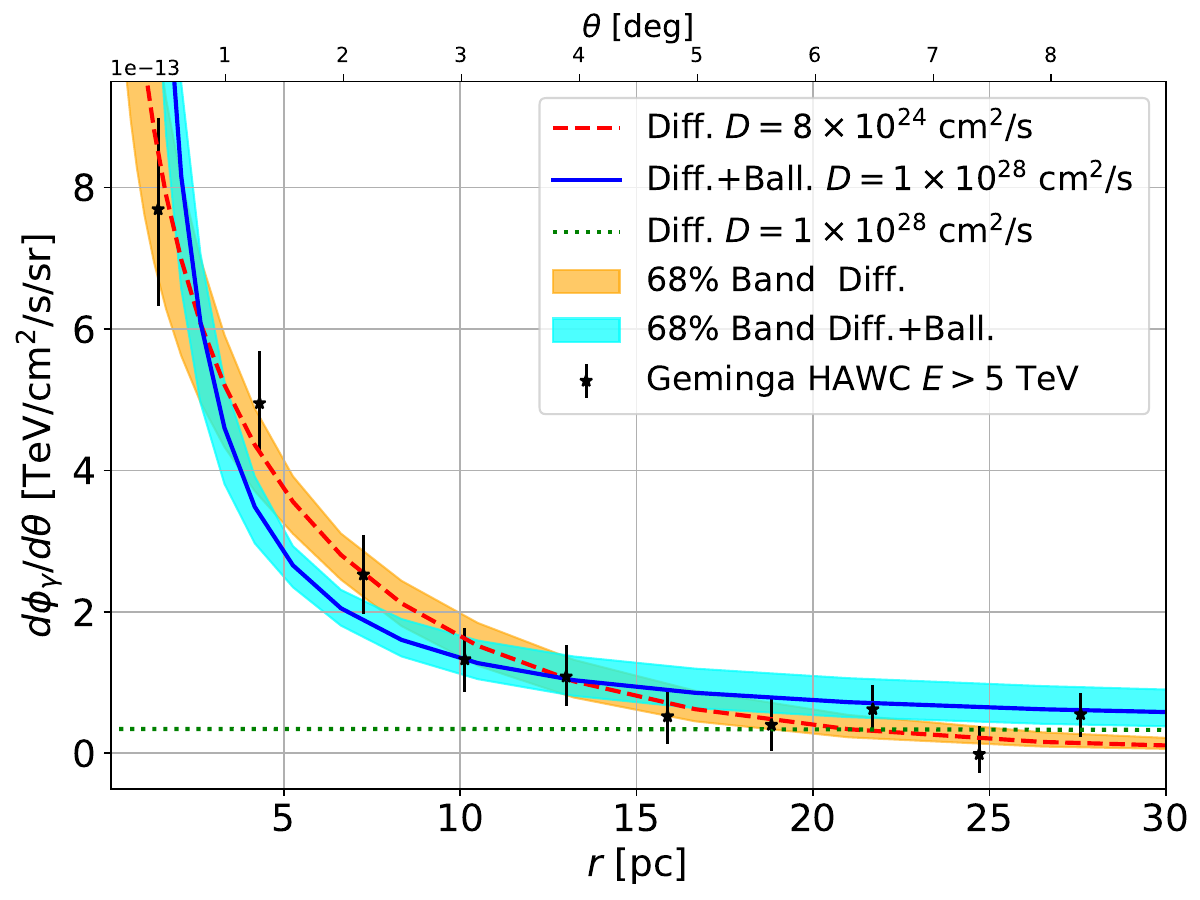}
\includegraphics[width=1.0\columnwidth]{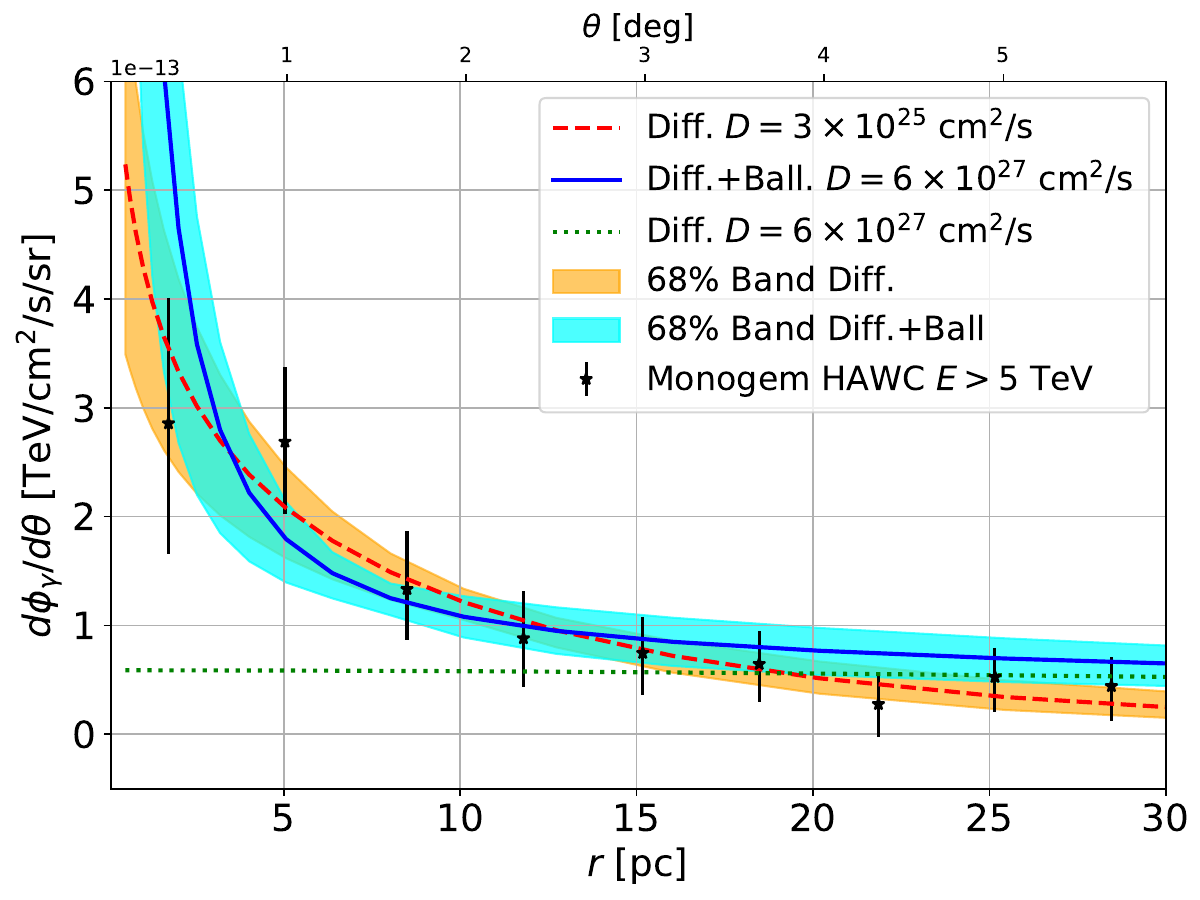}
\includegraphics[width=1.0\columnwidth]{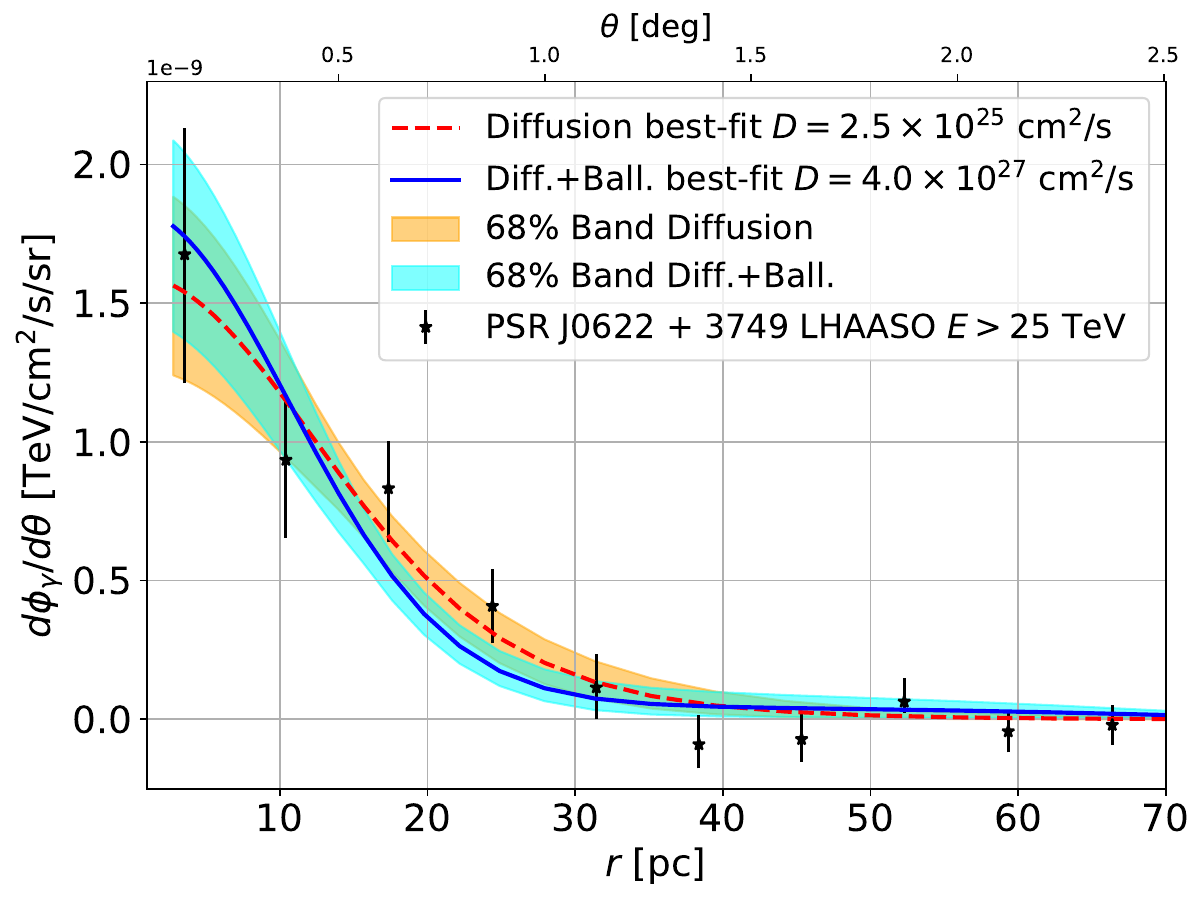}
\caption{
Fit to the HAWC data for Geminga (top left panel) and Monogem (top right panel), and to the PSR J0622+3749 (bottom panel) in the diffusive regime (red dotted line) and in the combined diffusive and ballistic model (blue solid line and cyan band). We show here the case where the distance of the Geminga and Monogem pulsars are 0.19 and 0.288 kpc, respectively. }
\label{fig:Hawc-geminga}
\end{figure*}

In Fig.~\ref{fig:Hawc-geminga} we show the spatial distribution of the $\gamma$-ray flux for our best fit model to the HAWC and LHAASO surface brightness data, both in the case of the diffusive-only (relevant for small values of the diffusion coefficient) and of the complete model, i.e.~diffusive plus ballistic. 
When the transition from ballistic to diffusive regime is properly taken into account, a good fit to the data can be achieved without invoking a very small diffusion coefficient.
In fact for $D_0 \gtrsim 8\times 10^{27} \rm cm^2/s$ the ballistic regime provides a surface brightness with a shape that goes as $\approx 1/r$ that fits well the data. 
Instead, when the ballistic-diffusion transition is ignored and the diffusive regime is applied even for very recent emission, one is forced to invoke a small diffusion coefficient to explain the spatial profile.
This can be seen in Fig.~\ref{fig:Hawc-geminga}, where the green dotted curves show the $\gamma-$ray surface brightness that one would get if pure diffusion is applied with a typical Galactic diffusion coefficient. Without the inclusion of the ballistic-diffusion transition, one would  inevitably be lead to the conclusion that a typical Galactic diffusion coefficient cannot reproduce the data.
In the case of $D_0$ of the order of $\sim 10^{25}$ cm$^2$/s, the exponential term $\exp(-r^2/r_d^2)$ that appears in the diffusive solution (see Eq.~\ref{eq:fe-full}) starts to be relevant for distances $r \gtrsim r_d \sim \sqrt{4\,D\,t_{\rm{loss}}}$ \citep{Recchia-2019PhRvD..99j3022R}, allows to fit the spatial profile. 
The $\gamma$ rays observed at energies $5-50$ TeV are produced by $e^{\pm}$ of energies between $20-200$ TeV. For these leptons and for $D_0 \sim  10^{25} \rm cm^2/s$ the scale at which the exponential factor becomes relevant is thus $r > $ a few pc, that is exactly the scale at which the $\gamma-$ray data decreases with the distance from the source. This also explains  why the best-fit $D_0$ decreases with smaller values of $B$. In fact, an increase of the loss time has to be compensated with a  decreased $D_0$ in order to get the same spatial extension.
This illustrates how the estimation of $D_0$ is sensitive to the chosen parameters in the  low-$D_0$ scenario, a problem which is much less prominent in the scenario proposed here.\\

During the review process of our paper Ref.~\cite{Bao-2021arXiv210707395B} appeared on Arxiv. The authors question the validity of our results, based on the high required efficiency for the Geminga pulsar, and for PSR J0622+3749 for the much worse fit to the $\gamma-$ray surface brightness they obtain with the ballistic with respect to the purely diffusive scenario. 
In their analysis they do not include the correction for the angular distribution of particles in the quasi-ballistic regime, namely the $M(\mu)$ function (see Eq.~\ref{eq:Mmu} and \ref{eq:Fe-full}). Moreover, they adopt a slope for the diffusion coefficient of $\delta = 1/3$.

As for Geminga, they use a power-law injection spectrum with slope $\alpha = 1.0$ and cut-off energy $E_c = 150$ TeV. With such setup they obtain a best-fit $D_0 = 2\times 10^{29}\, \rm cm^2/s$ and an efficiency  $\eta \sim 380\%$.
Using the same injection spectrum and the same slope of the diffusion coefficient we obtain a best-fit $D_0 \approx 5\times 10^{28}\, \rm cm^2/s$ and $\eta = 140\%$. The difference can be explained with the effect of $M(\mu)$ on the $\gamma-$ray morphology, which tends to make the radial profile of the surface brightness steeper than the data at high values of the diffusion coefficient. This tends to favor smaller values of $D_0$ when $M(\mu)$ is taken into account, which reflects also in a smaller value of the required efficiency. Notice also the very relevant impact of the chosen injection spectrum on the efficiency. Indeed, when the slope is changed from $\alpha = 1.0$ to $\alpha = 1.5$ (our benchmark case), the best-fit efficiency increases from $\sim 140\%$ to $\sim 180-200\%$.

As for the case of PSR  J0622+3749, the difference between our  results and that of  Ref.~\cite{Bao-2021arXiv210707395B} can be explained with considerations similar to the case of Geminga. Moreover, when applying the convolution with the PSF of LHAASO, we get a shape of the surface brightness similar to the PSF shape ( as it should be since this process makes the emission almost point like) that matches well the LHAASO data, as shown in Fig.~\ref{fig:Hawc-geminga}.  Ref.~\cite{Bao-2021arXiv210707395B} instead obtains a much flatter surface brightness in the case of quasi-ballistic propagation compared to the pure diffusion case when applying the PSF convolution, which is difficult to explain even considering that they use a size of the PSF of $0.45^{\circ}$ while we use $0.38^{\circ}$. This choice should play a minor role in the difference of the results.

As illustrated above, overall we get an efficiency below 100\% both for PSR J0622+3749 and for Monogem. 
Considering all the uncertainties discussed in this sections and that high efficiencies are indeed expected in pulsars, our model appears to be compatible with current data.

\section{Summary}

In this paper we demonstrate that the propagation of $e^{\pm}$, injected by pulsars, is dominated by the quasi-ballistic regime up to distances from the source of the order of $\lambda_c$, which is about 30 pc at multi-TeV energies. When the transition between the quasi-ballistic and diffusive regime is taken into account, it is possible to fit the HAWC data for Geminga and Monogem and the LHAASO data for PSR J0622+3749  with typical values of the diffusion coefficient used to fit CR data \cite{Genolini:2019ewc,DiMauro:2021qcf}, without invoking a hardly justifiable suppression. 
The currently available data for there three sources do not allow to reliably discriminate between the two scenarios because they are particularly different within $0.5^{\circ}$ from the source where the effect of the instrument PSF and the assumed size of the bow shock is very relevant. The future detection of halos around middle-age pulsars with $T>50$ kyrs, i.e.~pulsars not confined in the parent SNR  \cite{Blasi-2011ASSP...21..624B}, and with a distance of 1-5 kpc would provide us important hint on which between the quasi-ballistic or the pure-diffusive scenario takes place. 

Indeed, for such pulsars a small diffusion coefficient $D_0\sim 10^{25}$ cm$^2$/s would inevitably lead to a very small angular size, while for values of the  diffusion coefficient similar to the Galactic average $D_0\sim 10^{28}$ cm$^2$/s, the overall extension is expected to be much larger, as due to the diffusive part of the full transport solution. 
Therefore, in case of an inhibited diffusion these sources should be detected by HAWC and LHAASO \cite{LHAASO:2021zta}  as very compact while with the Galactic average diffusion they would look like as a bright spot, associate to the quasi-ballistic part, surrounded by an extended faint emission due to the diffusive part. 

In a followup paper we will discuss the different spacial morphology for the low and high-diffusion scenarios in case of sources with different ages and distances. In this forthcoming paper we will also report prospects for detection of $\gamma$-ray halos with current and future experiments.

\bibliographystyle{apsrev4-1}
\bibliography{biblio}

\section*{Acknowledgments}
MDM research is supported by Fellini - Fellowship for Innovation at INFN, funded by the European Union’s Horizon 2020 research programme under the Marie Skłodowska-Curie Cofund Action, grant agreement no.~754496.
SG acknowledges support from the Agence Nationale de la Recherche (grant ANR- 17-CE31-0014), and from the Observatory of Paris (Action F\'ed\'eratrice CTA). 
The work of FD and SR is partially supported by 
the {\sc Departments of Excellence} grant awarded by the Italian Ministry of Education,
University and Research ({\sc Miur}), the 
Research grant {\sl The Dark Universe: A Synergic Multimessenger Approach}, No.
2017X7X85K funded by the {\sc Miur} and by the 
Research grant {\sc TAsP} (Theoretical Astroparticle Physics) funded by Istituto
Nazionale di Fisica Nucleare. 

\clearpage

\end{document}